\documentclass[12pt,letterpaper]{article}

\usepackage{graphicx}

\begin{document}

\centering

\textbf{First-principles Study of Electronic and Dielectric Properties 
of ZrO$_2$ and HfO$_2$} 

\raggedright

\vspace{12pt}
Xinyuan Zhao and David Vanderbilt \\
Department of Physics and Astronomy, \\
Rutgers University, \\
Piscataway, NJ 08854-8019

\vspace{20pt}
\textbf{ABSTRACT}

\vspace{12pt}
\hspace{0.3in}
Using density-functional theory with ultrasoft pseudopotentials,
we previously investigated the structural and electronic
properties of the low-pressure (cubic, tetragonal, and monoclinic)
phases of ZrO$_2$ and HfO$_2$, in order to elucidate
phonon modes, Born effective charge tensors, and especially the
lattice dielectric response in these phases. We now extend
this previous work by carrying out similar calculations on the
two high-pressure orthorhombic phases, and by providing
density-of-states and band-gap information on all
polymorphs. Our results show that the electronic structures and
dielectric responses are strongly phase-dependent.  In particular,
the monoclinic phases of ZrO$_2$ and HfO$_2$ are found to
have a strongly anisotropic dielectric tensor and a rather
small orientational average ($\bar{\epsilon_0}$) compared to
the two other low-pressure phases. Our calculations show that
$\bar{\epsilon_0}$ is even smaller in the orthorhombic phases.

\vspace{26pt}
\textbf{INTRODUCTION}

\vspace{12pt}
\hspace{0.3in}
Dielectrics comprise a class of materials of great technical
importance. The most widely-used dielectric in electronics is
probably SiO$_2$, which has been persistently exploited in the computer
industry for decades due to the excellent quality of the Si/SiO$_2$
interface. The dimensions of the gate-dielectric layer (currently SiO$_2$)
have been rapidly reduced to the nanometer scale in order to enhance
the speed of CMOS-related devices. Such reductions, however,
have uncovered severe fundamental problems that imposed a
physical limit on the use of SiO$_2$ films as gate dielectrics in the
current Si/SiO$_2$ semiconductor technology.  A great deal of effort has
been directed to finding a way to overcome these limits. One
particular direction is to use high-$K$ metal oxides with certain
desired properties [1] to replace SiO$_2$ as the gate dielectric,
since the higher dielectric constants of the oxides will allow one to
make a physically thicker film while at the same time maintaining
(or even increasing) the gate capacitance.

\vspace{12pt}
\hspace{0.3in}
ZrO$_2$, HfO$_2$ and their structure-modified derivatives (e.g., Zr
and Hf silicates) have emerged as important candidates for this
purpose because they have much higher dielectric constants than
SiO$_2$ ($\sim$20 compared with 3.5 for SiO$_2$) and because
of their thermodynamic stability in contact with Si. Experiments
have demonstrated their promise for this purpose (see Ref.~[1]
for a general review).  It would be very desirable, therefore,
to conduct a systematic theoretical study of the structural,
electronic, and dielectric properties of ZrO$_2$ and HfO$_2$, and
of the dependence of these properties on the choice of crystalline
polymorphs or possible modified structures.  In two previous
works, we studied the structural and vibrational properties of the
three low-pressure phases (cubic, tetragonal, and monoclinic) of
ZrO$_2$ [3] and HfO$_2$ [4], especially the zone-centered phonon
modes and the related dielectric properties.  These previous works
are extended here in two ways. First, we provide density-of-states
(DOS) information
on all polymorphs of ZrO$_2$ and HfO$_2$, which may be helpful
in the interpretations of experiments on these materials.
Second, we extend our previous calculations of dielectric
properties to include the high-pressure orthorhombic phases
using the ABINIT package [4].  We have also been engaged in
simulating amorphous ZrO$_2$ ($a$-ZrO$_2$) using {\em ab-initio}
molecular dynamics as implemented in the VASP [5] package, in a
``melt-and-quench'' fashion. The resulting amorphous ZrO$_2$
structures were analyzed in order to understand their local
bonding geometry and other structural properties, and then used
as inputs for a first-principles investigation of the vibrational
and dielectric properties of the amorphous ZrO$_2$ systems using
ABINIT. The results will be reported elsewhere.

\vspace{20pt}
\textbf{CRYSTAL STRUCTURES}

\vspace{12pt}
\hspace{0.3in}
ZrO$_2$ (zirconia) is surprising similar to HfO$_2$ (hafnia) in
many physical and chemical properties. It is well
established that both oxides exhibit three crystalline phases at
ambient pressure, i.e., monoclinic, tetragonal and cubic, as
temperature is increased. With increasing pressure, phase transitions
involving two orthorhombic phases have been reported, although
the exact structures of these high-pressure phases has not been
established unambiguously ([6,7], and references therein). The first
orthorhombic structure used in our simulation (conventionally labeled
as Ortho-I) has space group $Pbca$; the other orthorhombic
phase (Ortho-II) has space group $Pnma$.  We study these phases
at zero pressure, where they are apparently metastable.

\vspace{12pt}
\hspace{0.3in}
The cubic phase has the fluorite (CaF$_2$) structure, with 3 atoms
(one chemical formula) per primitive unit cell. The tetragonal
structure is a slight distortion of the cubic phase obtained by shifting
alternating pairs of oxygen atoms up and down along $\hat{z}$,
and the number of atoms per primitive cell is doubled. The
monoclinic phase, which has 12 atoms (four formulas units) per
primitive cell, has a more complex structure and lower symmetry. In
our calculations of the DOS functions, a 12-atom cell has been used
for all of these phases, as in Refs.~[2] and [3], in order to
facilitate comparison.  The monoclinic phase has two non-equivalent
oxygen atoms, which evidently will have different site-projected
DOS functions. Among the eight oxygens
in the unit cell, four are threefold coordinated (denoted
as O$_1$), while the other four (labeled as O$_2$) are
fourfold coordinated. Each threefold oxygen is bonded to three
nearest-neighbor metal atoms in an almost planar configuration,
while each four-fold oxygen forms a distorted tetrahedron with
its four nearest metal neighbors. For the cubic and tetragonal
phases, all oxygen atoms are equivalent, although we still use
the notation O$_1$ and O$_2$ for convenience.  Detailed discussion
of the structures of these phases can be found in Refs.~[2] and [3].

\vspace{12pt}
\hspace{0.3in}
A primitive unit cell of the Ortho-I phase (for both ZrO$_2$ and
HfO$_2$) consists of 24 atoms, while the Ortho-II phase has 12
atoms per unit cell. We used the lattice parameters given in [6]
(Table III there) and [7] (Tables II and III there) as the starting
point for our ground-state calculations for Ortho-I and Ortho-II,
respectively. The Wyckoff coordinates define two non-equivalent
sets of oxygen atoms, labeled by O$_1$ and O$_2$.

\vspace{20pt}
\textbf{THEORETICAL METHODOLOGY}

\vspace{12pt}
\hspace{0.3in}
The DOS functions are computed using a plane-wave
implementation of density functional theory (DFT) in the
local-density approximation (LDA) as parametrized by Ceperley
and Alder [8], using ultra-soft pseudopotentials [9].  Although a
4$\times$4$\times$4 {\bf k}-point sampling was found to provide
sufficient precision in calculating total energies and forces,
we used a refined mesh of {\bf k}-points and higher 36-Ry
cut-off to obtain smoother and more accurate DOS
functions. The {\bf k}-point meshes used in the DOS calculations
are 20$\times$20$\times$20 for cubic and tetragonal,
10 $\times$10$\times$10 for monoclinic, 12$\times$12$\times$12 for
Ortho-I, and 16$\times$16$\times$16 for Ortho-II phases. In calculating
the site-projected partial DOS, spherical radii of 0.9\,\AA\, and
1.25\,\AA\, are used for the metal ions (Zr or Hf) and oxygen ions
respectively. For cubic and tetragonal phases, the larger
12-atom unit cell was used in the calculations, so the
DOS values should be reduced accordingly if the DOS per primitive cell
is desired. Further details can be found in [2] and [3].

\vspace{12pt}
\hspace{0.3in}
The dielectric response of orthorhombic phases of ZrO$_2$
was calculated by specialized linear-response techniques as
implemented in ABINIT [4], taking into account the coupling between
phonons and the homogeneous electric field. A 4$\times$4$\times$4
{\bf k}-point mesh, Troullier-Matins type pseudopotentials
[10] (4 and 6 valence electrons for Zr and O respectively), and
an energy cutoff of 35.0 Hartree provide satisfactory
convergence.

\begin{table}
\Roman{table}
\caption{Structural parameters calculated for the orthorhombic
phases of ZrO$_2$ and HfO$_2$.  Lattice constants
$a$, $b$ and $c$, in \AA\, are given in the first row.
Triplets of coordinates indicate the Wyckoff positions of the
various atoms.}
\medskip
\begin{tabular}{|l|c|c|} \hline
\textbf{Phase} & \textbf{ZrO$_2$} & \textbf{HfO$_2$} \\
\hline
       & $a$=9.958, $b$=5.224, $c$=5.006   & $a$=9.837, $b$=5.129, $c$=4.954 \\
Ortho-I & Zr (0.885, 0.035, 0.254)    &  Hf (0.885, 0.035, 0.254)\\
       & O$_1$ (0.792, 0.379, 0.133) & O$_1$ (0.791, 0.376, 0.128)\\
       & O$_2$ (0.976, 0.739, 0.497) & O$_2$ (0.977, 0.739, 0.497)\\
\hline
       &  $a$=6.438, $b$=5.521, $c$=3.273  &   $a$=6.353, $b$=5.435, $c$=3.240 \\
Ortho-II& Zr (0.884, 0.256, 0.250) &  Hf (0.885, 0.256, 0.250) \\
       & O$_1$ (0.164, 0.027, 0.250) &  O$_1$ (0.163, 0.025, 0.250) \\
       & O$_2$ (0.572, 0.142, 0.250) &  O$_2$ (0.573, 0.141, 0.250) \\
\hline
\end{tabular}
\label{table:phases}
\end{table}

\vspace{20pt}
\textbf{RESULTS}

\begin{figure}
\rotatebox{270}{
\includegraphics[scale=0.32]{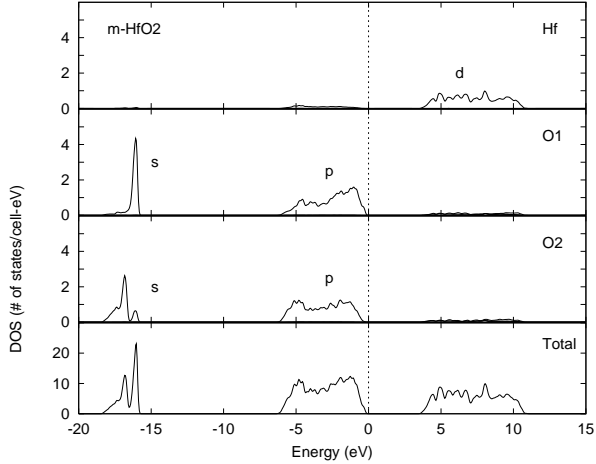}
}
\caption{Density of states (DOS) for monoclinic HfO$_2$. Upper panels
show the orbital-projected partial DOS for Hf, O$_1$, and O$_2$,
respectively. Lower panel shows the total DOS.
Vertical dashed line indicates the valence-band maximum.}
\label{fig:mono_pdos}
\end{figure}

\vspace{12pt}
\hspace{0.3in}
In Table~\ref{table:phases} we present our calculated lattice
parameters for the two orthorhombic phases, which agree well
with the previous theoretical calculations [6,7].
Because the computed DOS functions for HfO$_2$ and ZrO$_2$ are
so similar, we only present those for HfO$_2$ here, shown in Figs.~(1)-(3).
Some comparisons are given later in
Table~\ref{table:diel} for both oxides.
Figure \ref{fig:mono_pdos} shows the total DOS (lowest panel) and
the site-projected partial DOS (upper three panels) for monoclinic
HfO$_2$.  We find this material to be an insulator with a
band gap of 3.5\,eV, in good agreement with previous
theoretical results (3.48\,eV in [11] and 3.6\,eV in [12]).
The bands in the energy range from $-$20\,eV to $-$15\,eV are
mostly due to O 2$s$ orbitals. The next bands, spanning between
$-$6\,eV and 0, are mainly O 2$p$ in character, and Hf 5$d$
orbitals dominate the unoccupied conduction bands ranging
from 3\,eV up to 11\,eV. The two non-equivalent oxygen sites have
noticeably different DOS features, as expected from their different
surroundings.

\begin{figure}[b!]
\rotatebox{270}{
\includegraphics[scale=0.65]{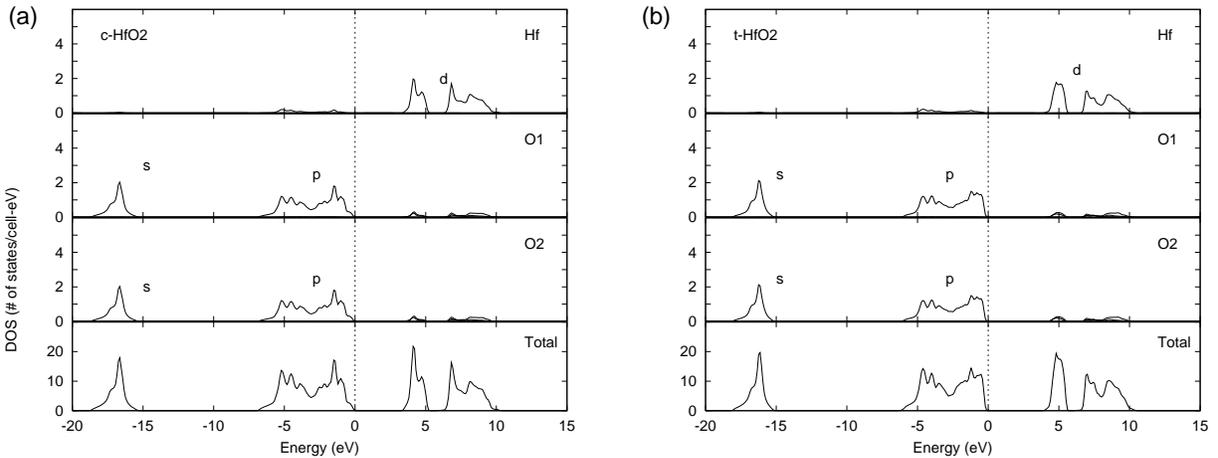}
}
\caption{Same as in Fig.~1, but now for cubic (a) and tetragonal (b)
phases of HfO$_2$.\qquad}
\label{fig:ct_pdos}
\end{figure}

\begin{figure}
\rotatebox{270}{
\includegraphics[scale=0.65]{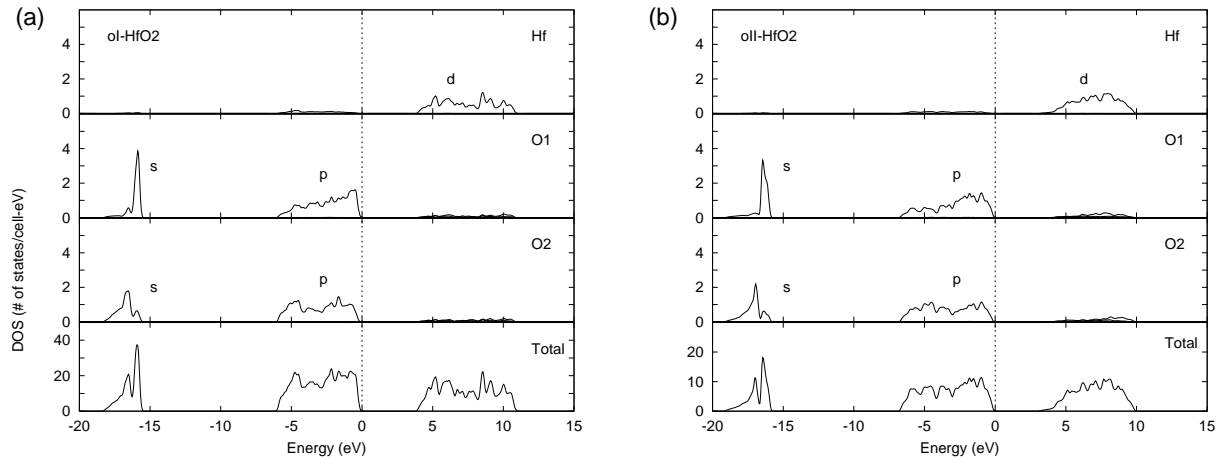}
}
\caption{Same as Fig.~1, but now for the two orthorhombic phases of
HfO$_2$.\qquad\qquad\qquad\quad}
\label{fig:o_pdos}
\end{figure}

\vspace{12pt}
\hspace{0.3in}
In Fig.~\ref{fig:ct_pdos}, we provide the DOS functions calculated
for cubic and tetragonal HfO$_2$. It can be readily seen
that these two phases have rather similar DOS features. Cubic
HfO$_2$ has a band gap of 3.2\,eV, while for tetragonal HfO$_2$
it is 3.8\,eV. Ref.~[11] gives 3.4\,eV and 3.77\,eV for the cubic
and tetragonal phases, respectively. Again, both theoretical
works agree quite well. The most significant difference between
these two phases and the monoclinic phase is that the Hf $d$-like
states separate into two bands in the cubic and tetragonal
phases, but form a single band in the monoclinic phase. The DOS
functions for the two high-pressure orthorhombic phases are
presented in Fig.~\ref{fig:o_pdos}, from which one readily sees
that the oxygen atoms exhibit two quite different types of DOS,
as in the monoclinic phase. In fact, there is a surprising degree
of resemblance of the monoclinic DOS with that of the Ortho-I and
Ortho-II phases, especially the former.

\vspace{12pt}
\hspace{0.3in}
The DOS functions of the ZrO$_2$ polymorphs (not
shown) are qualitatively very similar to the corresponding ones
for HfO$_2$.  The most important differences are in the band gaps.
Because of LDA error, the absolute values of the band gaps are not
reliable, but trends should be meaningful.  As shown in
Table~\ref{table:diel}, we find that the
band gaps are systematically larger (by $\sim$0.5\,eV) for HfO$_2$
than for ZrO$_2$, and that variations of crystal structure
can lead to band gap changes of order 1\,eV.
Also tabulated in Table~\ref{table:diel} are our calculated
dielectric constants (orientational average
$\bar{\epsilon_0}$) for ZrO$_2$ and HfO$_2$ polymorphs. For
the low-pressure cubic, tetragonal, and monoclinic phases,
the dielectric constants are quoted from [2] and [3], where
the full dielectric tensors are given. For the Ortho-I and
II phases of ZrO$_2$, the linear-response scheme [4] is utilized in
the calculation of the static dielectric tensors, which includes
the lattice contributions as well as the electronic screening.
In orthorhombic ZrO$_2$, $\bar{\epsilon_0}$ becomes even smaller,
approximately 20 and 18 for the Ortho-I and II phases, respectively.
The dielectric tensors for the orthogrombic phases are diagonal, with
elements (22.6, 18.1, 19.6) and (18.8, 18.9, and 17.8) for Ortho-I and
Ortho-II respectively.  The average principal values of the
Born effective charge tensors are about 5.0 and 2.5 for
Zr and O atoms respectively for both orthorhombic phases.

\begin{table}
\caption{Band gaps ($E_g$) and orientationally averaged dielectric
constants ($\bar{\epsilon_0}$) calculated for crystalline
phases of ZrO$_2$ and HfO$_2$. $E_g$ values in parentheses for HfO$_2$ 
are from the theory of Ref.~[11].}
\medskip
\begin{tabular}{|l|c|c|c|c|} \hline
\textbf{Phase} & \multicolumn{2}{c|}{\textbf{ZrO$_2$}} 
               & \multicolumn{2}{c|}{\textbf{HfO$_2$}} \\
\hline
& $E_g$ (eV) & ~~~~$\bar{\epsilon_0}$~~~~ & $E_g$ (eV)
 & ~~~~$\bar{\epsilon_0}$~~~~ \\
\hline
Cubic      & 2.63 & 37 & 3.15 (3.40) & 29 \\
Tetragonal & 3.31 & 38 & 3.84 (3.77) & 70 \\
Monoclinic & 2.98 & 20 & 3.45 (3.48) & 16--18 \\
\hline
Ortho-I     & 3.06 & 20.1 & 3.75 & - \\
Ortho-II    & 2.29 & 18.5 & 2.94 & - \\
\hline
\end{tabular}
\label{table:diel}
\end{table}

\vspace{20pt}
\textbf{CONCLUSIONS}

\vspace{12pt}
\hspace{0.3in}
We studied the structural, electronic, vibrational
and dielectric properties for nearly all the currently
well-recognized phases of ZrO$_2$ and HfO$_2$, using
advanced first-principles techniques, with a special focus on
the dielectric properties. It is found that the crystalline
structure can have a substantial effect on the band gaps and dielectric
constants.  Our results show that the dielectric response depends
strongly on the crystal phase, and can span a wide range of values
($\bar{\epsilon_0}$ = 18 -- 40 for ZrO$_2$). Monoclinic ZrO$_2$
(or HfO$_2$) has a rather anisotropic dielectric tensor and a
smaller averaged dielectric constant than the tetragonal and cubic
phase. The $\bar{\epsilon_0}$ of the Ortho-I and II phases appear
even smaller. We hope that these theoretical results will shed
light on the potential for ZrO$_2$, HfO$_2$ or their derivatives
to replace SiO$_2$ as the gate dielectric in CMOS technology,
and will provide hints for manipulating these oxides to serve
this purpose.

\vspace{20pt}
\textbf{ACKNOWLEDGMENTS}

\vspace{12pt}
\hspace{0.3in}
This work has been supported by NSF Grant DMR-9981193. We wish
to thank E. Garfunkel and S. Sayan for useful discussions. One
of authors (X.Z.) thanks G.-M.\ Rignanese for assistance with ABINIT.
 
\vspace{20pt}
\textbf{REFERENCES}

\begin{enumerate}

\item G. D. Wilk, R. M. Wallace, and J. M. Anthony, J. Appl. Phys. 
{\bf 89}, 5243 (2001).
\vspace{-0.13in}

\item X. Zhao and D. Vanderbilt, Phys. 
Rev. B {\bf 65}, 75105 (2002).
\vspace{-0.13in}

\item X. Zhao and D. Vanderbilt,  Phys. 
Rev. B {\bf 65} 233106 (2002).
\vspace{-0.13in}

\item X. Gonze et al., Mater. Sci. {\bf 25}, 478 (2002). 
\vspace{-0.13in}

\item G. Kresse and J. Hafner, Phys. Rev. B {\bf 47}, R558 (1993);
{\bf 54}, 11169 (1996).
\vspace{-0.13in}

\item J. K. Dewhurst and J. E. Lowther, Phys. 
Rev. B {\bf 57}, 741 (1998).
\vspace{-0.13in}

\item G. Jomard, T. Petit, A. Pasturel, L. Magaud, 
G. Kresse, and J. Hafner, Phys. Rev. B {\bf 59}, 4044 (1999).
\vspace{-0.13in}

\item D. M. Ceperley and B. J. Alder, Phys. Rev. Lett. {\bf
45}, 566 (1980).
\vspace{-0.13in}

\item D. Vanderbilt, Phys. Rev B {\bf 41}, 7892 (1990).
\vspace{-0.13in}

\item N. Troullier and J. L. Martins, Phys. Rev. B {\bf 43}, 
1993 (1991).
\vspace{-0.13in}

\item A. A. Demkov, Phys. Stat. Sol. B {\bf 226}, 57 (2001).
\vspace{-0.13in}

\item P. K. Boer and R. A. de Groot, J. Phys.:
Condens. Matter {\bf 10}, 10241 (1998).
\vspace{-0.13in}

\end{enumerate}

\newpage

\end{document}